\documentclass[aps,twocolumn,showpacs,showkeys,superscriptaddress]{revtex4-1}
\usepackage[T1]{fontenc}
\usepackage{amsfonts,amsmath,amssymb,amsbsy}
\usepackage{graphicx,color}
\usepackage{times}
\usepackage[colorlinks=true, linkcolor=blue, citecolor=red, urlcolor=magenta]{hyperref}
\let\mathbf=\boldsymbol

\def\blue#1{\textcolor{blue}{#1}}

\begin{document}

\title{{\Large Creation, transport and detection of imprinted magnetic solitons \\ stabilized by spin-polarized current}}

\author{R.\ P.\ Loreto}
\affiliation{Departamento de F\'{i}sica, Universidade Federal de Vi\c cosa, Vi\c cosa, 36570-900, Minas Gerais, Brazil}

\author{W.\ A.\ Moura-Melo}
\affiliation{Departamento de F\'{i}sica, Universidade Federal de Vi\c cosa, Vi\c cosa, 36570-900, Minas Gerais, Brazil}

\author{A.\ R.\ Pereira}
\affiliation{Departamento de F\'{i}sica, Universidade Federal de Vi\c cosa, Vi\c cosa, 36570-900, Minas Gerais, Brazil}

\author{X. Zhang}
\affiliation{School of Science and Engineering, The Chinese University of Hong Kong, Shenzhen 518172, China}

\author{Y. Zhou}
\affiliation{School of Science and Engineering, The Chinese University of Hong Kong, Shenzhen 518172, China}

\author{M. Ezawa}
\affiliation{Department of Applied Physics, University of Tokyo, Hongo 7-3-1, 113-8656, Japan}

\author{C.\ I.\ L.\ Araujo}
\email[E-mail:~]{dearaujo@ufv.br}
\affiliation{Departamento de F\'{i}sica, Universidade Federal de Vi\c cosa, Vi\c cosa, 36570-900, Minas Gerais, Brazil}

\begin{abstract}
With the recent proposition of skyrmion utilization in racetrack memories at room temperature, skyrmionics has become a very attractive field. However, for the stability of skyrmions, it is essential to incorporate the Dzyaloshinskii-Moriya interaction (DMI) and the out-of-plane magnetic field into the system. In this work, we explore a system without these interactions. First, we propose a controlled way for the creation of magnetic skyrmions and skyrmioniums imprinted on a ferromagnetic nanotrack via a nanopatterned nanodisk with the magnetic vortex state. Then we investigate the detachment of the imprinted spin textures from the underneath of the nanodisk, as well as its transport by the spin-transfer torque imposed by spin-polarized current pulses applied in the nanotrack. A prominent feature of the moving imprinted spin texture is that its topological number $Q$ is oscillating around the averaged value of $Q=0$ as if it is a resonant state between the skyrmions with $Q=\pm1$ and the bubble with $Q=0$. We may call it a resonant magnetic soliton (RMS). A RMS moves along a straight line since it is free from the skyrmion Hall effect. In our studied device, the same electrodes are employed to realize the imprinted spin texture detachment and its transport. In addition, we have investigated the interaction between the RMS and a magnetic tunnel junction sensor, where the passing of the RMS in the nanotrack can be well detected. Our results would be useful for the development of novel spintronic devices based on moveable spin textures.
\end{abstract}

\date{\today}

\maketitle

\section{Introduction}
\label{se:introduction}

The magnetic skyrmion~\cite{skyrme,rossler}, a sort of soliton with protected structural stability assured by its topological configuration, was recently observed in chiral magnetic materials~\cite{bogdanov,bode,Muhlbauer,neubauer,pappas,Yu} with the Dzyaloshinskii-Moriya interaction (DMI)~\cite{dzy,moriya} at low temperatures. Moreover, experimental observations performed at room temperature, in materials with both perpendicular magnetic anisotropy (PMA) and DMI~\cite{YuR,buttner,hei}, together with the topological protection and fast transport~\cite{schulz,seki}, have pointed out skyrmions as the most prominent magnetic structures to be exploited for building magnetic storage devices, such as the racetrack memories~\cite{parkin1}. The magnetic domain wall-based racetrack memory is an established technique~\cite{tehrani,par,araujo,yang}. The merit of the skyrmion compared with the domain wall is that the skyrmion is not easy to be pinned by defects or impurities~\cite{iwasaki}. The manipulation of a single isolated skyrmion is a key for skyrmionics~\cite{iwasaki2,silva,sampaio,Yin,zhang}. The logic computing application and electric devices based on skyrmions make skyrmionics future promising spintronic devices~\cite{zhang9400,kang,xing1,zhangV}.

Recent experimental works have demonstrated new methods for the generation of single isolated skyrmion~\cite{woo}, skyrmions form on a nanotrack with PMA using protocol of out-of-plane magnetic saturation and successive magnetic field reversion. Another promising experimental method~\cite{jiang} is the transformation of skyrmions from chiral domain walls in a nanotrack with the geometrical constriction. However, in all these experimental works carried out so far, an out-of-plane magnetic field is required to maintain the stability of skyrmions.

\begin{figure}[t]
\centering
\includegraphics[width=0.50\textwidth]{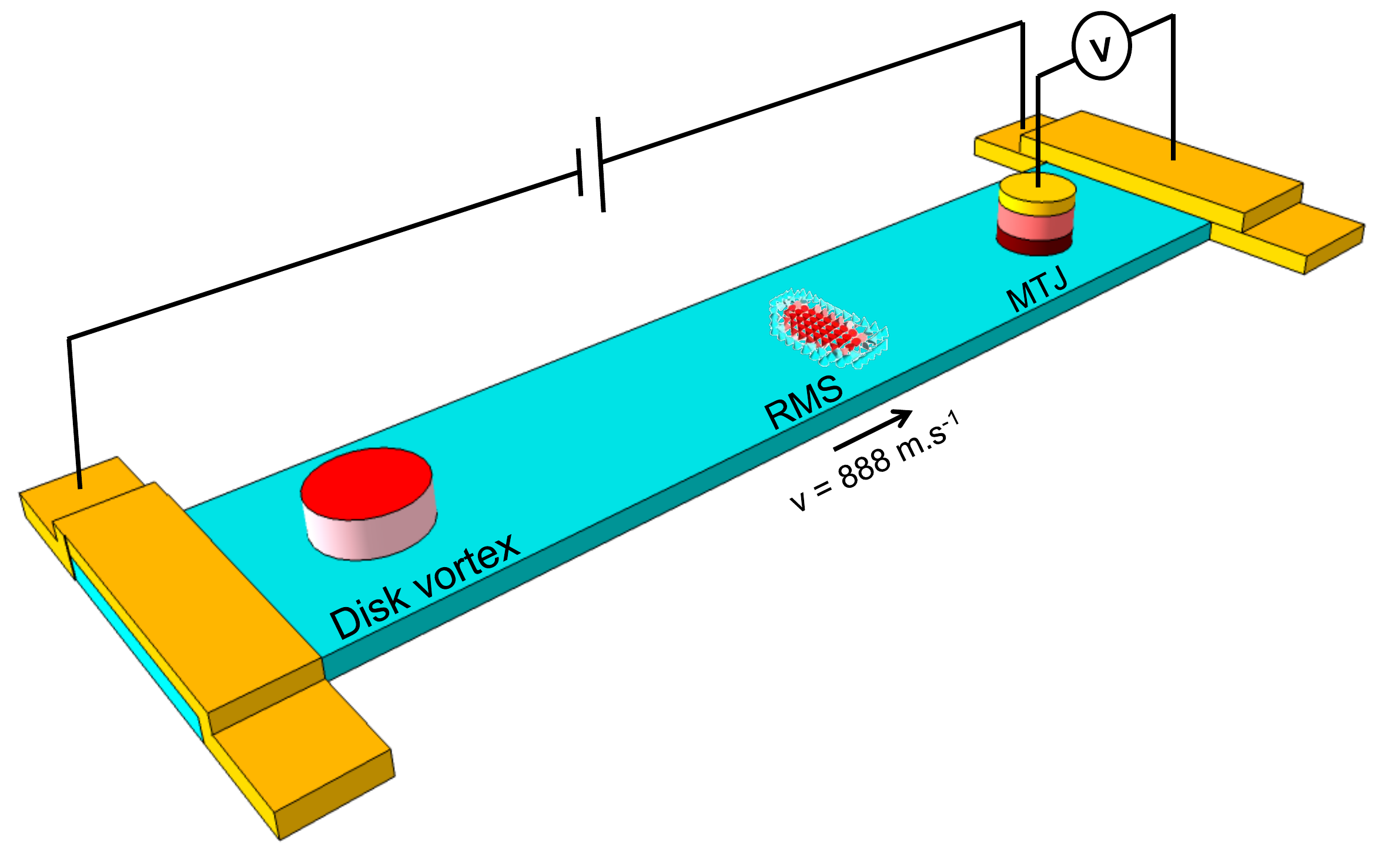}
\caption{(Color online.) Scheme of the proposed device where a Co nanodisk with the diameter of $120$ nm and the thickness of $20$ nm is patterned on a CoPt nanotrack with the length of $2$ $\mu$m, the width of $200$ nm, and the thickness of $8$ nm. The MTJ is composed by a $4$-nm-thick CoPt nanodisk separated from the nanotrack by a $2$-nm-thick tunnel barrier. The driving current is applied via the gold contacts on the nanotrack extremities, and the RMS passage is measured by tunnel magnetoresistance.}
\label{FIG1}
\end{figure}

\begin{figure*}[t]
\centering
\includegraphics[width=1.00\textwidth]{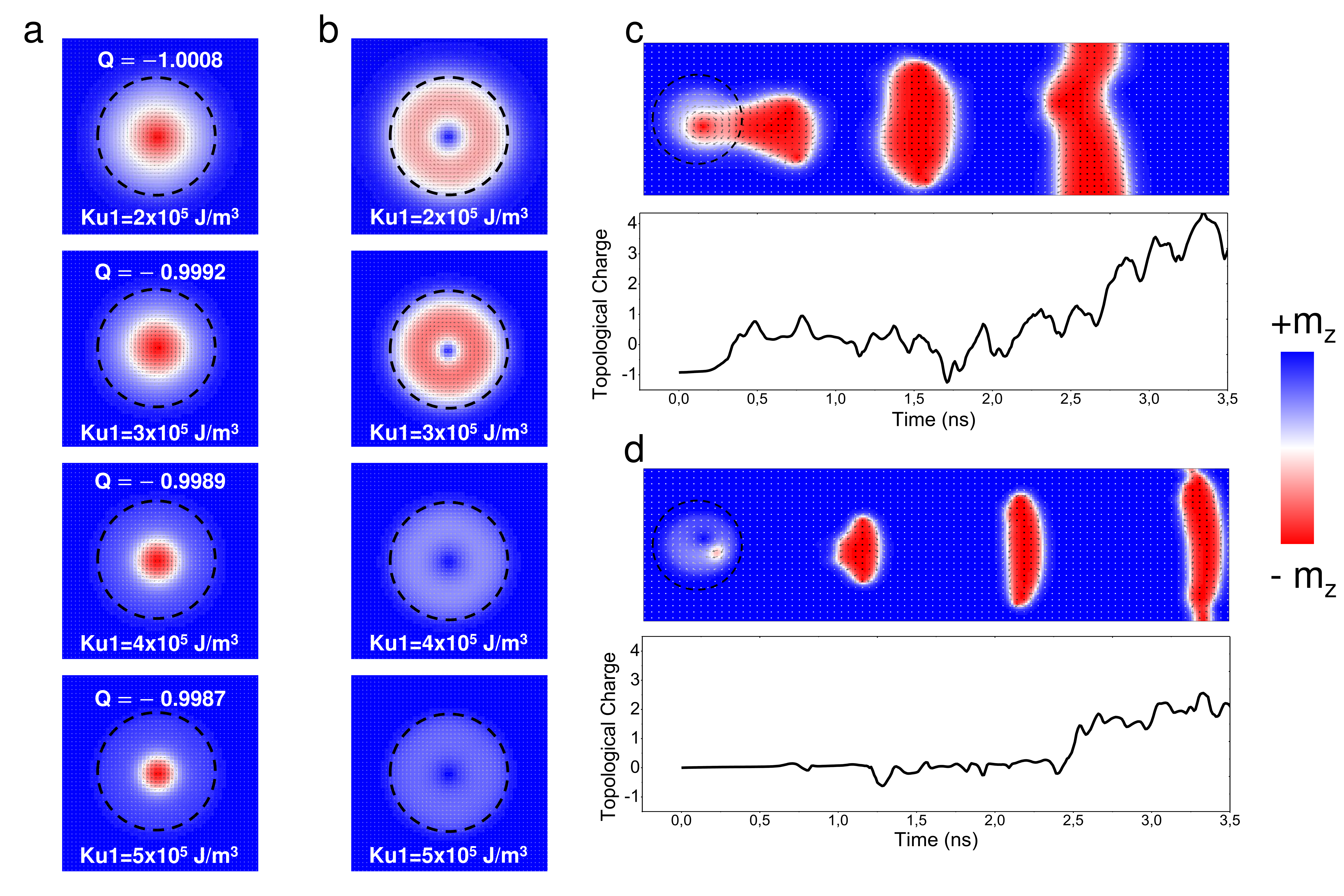}
\caption{(Color online.) (a) A skyrmion is imprinted on the nanotrack, after applying a small out-of-plane field in order to saturate the nanotrack magnetization. The corresponding topological number $Q$ is indicated. A dotted circle shows the size of the nanodisk. We have applied $B_z=0.15$ T. (b) A skyrmionium is obtained after applying a high out-of-plane field until the total system saturation. We have applied $B_z=0.5$ T. (c) The spin texture detachment from the underneath of the Co nanodisk driven by the current with a big bubble formation. The time evolution of $Q$ is also given. (d) A small bubble formation from the skyrmionium driven by the current as well as the Oersted field. A bubble converts to a domain wall pair when it touches the edges of the nanotrack. The time evolution of $Q$ is also given. The magnetization component $m_z$ is represented by the color scale, where white denotes the in-plane spins, while red and blue represent the opposite out-of-plane spin configurations.
}
\label{FIG2}
\end{figure*}

In this work, using full micromagnetic simulations, we propose to transport a new type of magnetic spin textures through a nanotrack where both the DMI and the out-of-plane magnetic field are absent. First, we create a skyrmion or a skyrmionium by the interaction between a magnetic nanotrack with PMA and a soft magnetic nanodisk with vortex state attached to it, as shown in Fig.~\ref{FIG1}. This method of creating a skyrmion imprinted by a nanopatterned nanodisk on a CoPt thin film has already been investigated by numerical simulations~\cite{sun} as well as experiments demonstrated at room temperature~\cite{miao,gilbert}. However, the possibility of detaching a skyrmion or a similar imprinted spin texture from the underneath of the nanodisk and its subsequent stability and transport properties are yet to be investigated. With the geometry proposed for a magnetic storage device, as illustrated in Fig.~\ref{FIG1}, we explore the capability of the nanodisk with the vortex state to imprint spin textures, such as the skyrmion, skyrmionium~\cite{finazzi,zhou2016}, bubble, and droplet~\cite{zhou2014, zhou2015}, on a nanotrack with different values of the PMA. Then, in order to promote the detachment of the imprinted spin texture (IST) from the underneath of the nanodisk and its evolution to a stable dynamic spin texture, traveling throughout the $2$-$\mu$m-long nanotrack, we develop a protocol of applying spin-polarized current pulses. A prominent feature of the moving IST is that its topological number $Q$ is oscillating around the averaged value of $Q=0$ as if it is a resonant state between the skyrmions with $Q=\pm1$ and the bubble with $Q=0$. This occurs because of the absence of the DMI and the out-of-plane magnetic field. Let us refer to such an IST as a resonant magnetic soliton (RMS). We also study the interaction between the moving RMS and a magnetic tunnel junction (MTJ) sensor, which will be utilized in the detection measurement of a moving RMS.

\begin{figure}[t]
\centering
\includegraphics[width=0.50\textwidth]{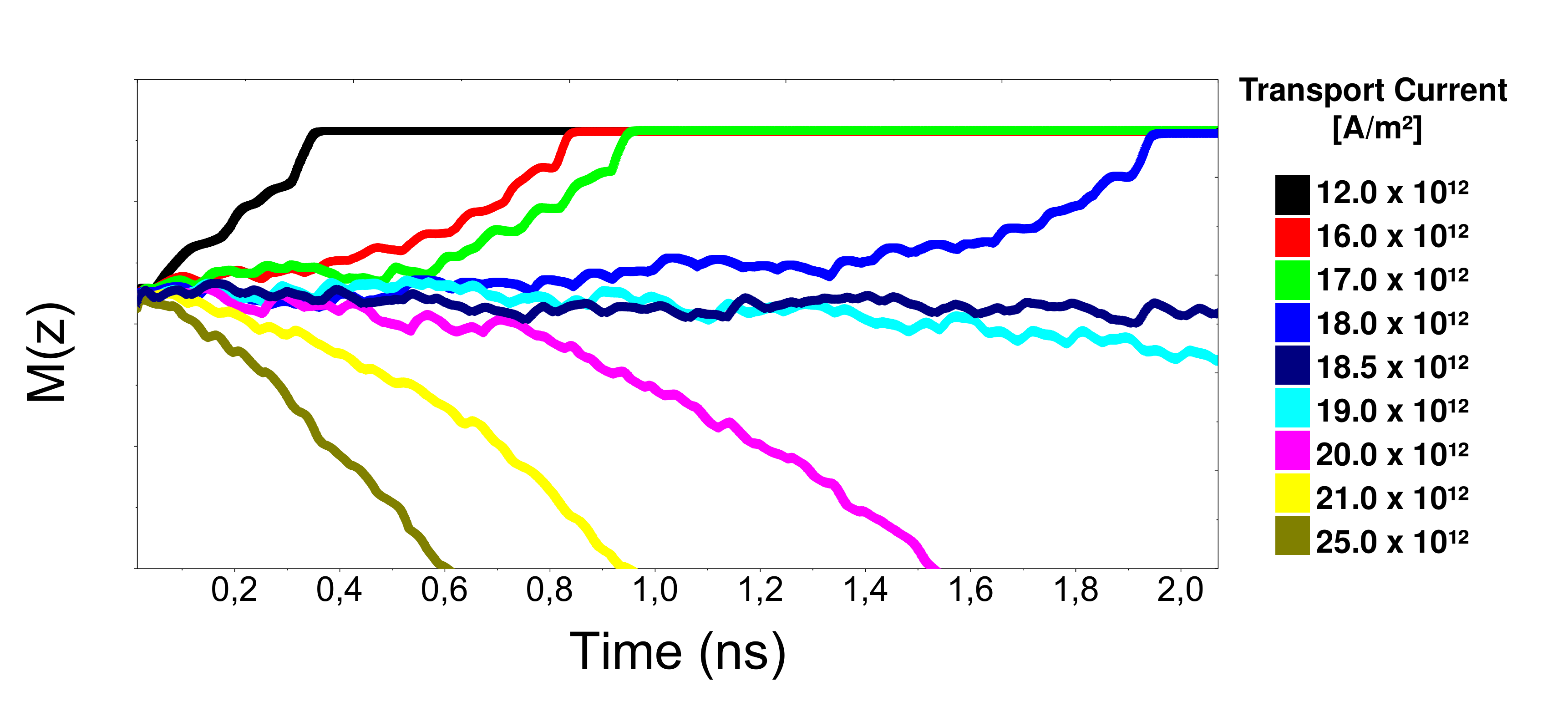}
\caption{(Color online.) Time evolution of the magnetization component $m_z$ for various magnitude of current. A magnetic bubble shrinks and disappears for the current weaker than $J_s=18.455\times 10^{8}$ A cm$^{-2}$. On the other hand, a magnetic bubble expands and transformed into a domain wall pair for the current stronger than $J_s$.}
\label{FIG3}
\end{figure}

\begin{figure}[t]
\centering
\includegraphics[width=0.50\textwidth]{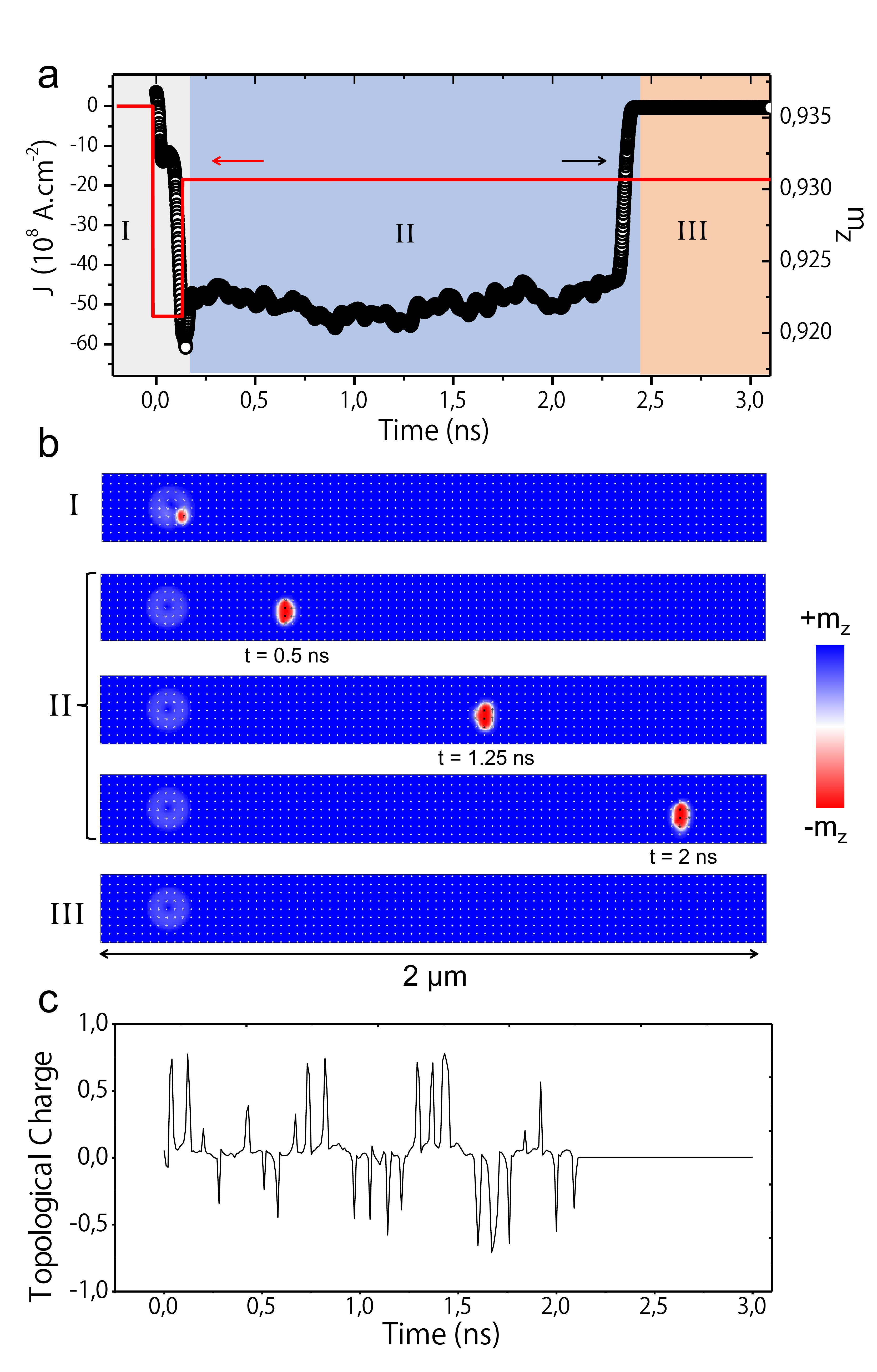}
\caption{(Color online.) (a) Time evolution of $m_z$ and $J$ during the creation and transport of the RMS. (b) Top-views of the magnetization in the nanotrack at selected times showing the creation of the RMS (I), the transport of the imprinted RMS (II), and the nanotrack where the RMS is expelled (III). (c) Topological number $Q$ of the nanotrack as a function of time. We have used $K_{\text{u1}}=5\times 10^5$ J m$^{-3}$}
\label{FIG4}
\end{figure}

\section{Methods}
The numeric simulations of the space- and time-dependent dynamics in ferromagnets, for energy minimization and transitions between spin configurations, are carried out based on the Landau-Lifshitz-Gilbert (LLG) equation augmented with the spin-transfer torque, given by
\begin{equation}
\frac{\partial\boldsymbol{M}}{\partial t}=
-\gamma\boldsymbol{M}\times\boldsymbol{H}_\text{eff}
+\frac{\alpha}{M_\text{S}}\boldsymbol{M}\times\frac{\partial\boldsymbol{M}}{\partial t}
+\frac{pa^3}{2eM_\text{S}}(\boldsymbol{J}(\boldsymbol{r})\cdot\triangledown)\boldsymbol{M}),
\label{eq:LLG}
\end{equation}
where $\gamma$ is the gyromagnetic ratio, $M_\text{S}$ the saturation magnetization, $\alpha$ is the Gilbert damping coefficient, $p$ is the polarization ratio of the electric current, $a$ is the lattice constant and $\boldsymbol{H}_\text{eff}$ is the effective field, which is composed by the magnetostatic field, the external magnetic field, the magnetocrystalline anisotropy, the Heisenberg exchange interaction, and the dipolar interaction. The third term on the right-hand side of Eq.~(\ref{eq:LLG}) is related to the adiabatic spin-transfer torque provided by the spin-polarized current $\boldsymbol{J}(\boldsymbol{r})$ in the nanotrack. For the iterations, we have utilized the GPU-accelerated micromagnetic simulator MuMax$^{3}$~\cite{mumax}.

The finite difference discretization used by MuMax$^{3}$ to follow Eq.~(\ref{eq:LLG}) is performed in a space of a two-dimensional or three-dimensional grid of orthorhombic cells. Here, we have utilized a cubic cell of $4$ nm $\times$ $4$ nm $\times$ $4$ nm for the iterations. Magnetic parameters for Co and CoPt are adopted from Ref.~\onlinecite{sun}. The diameter and thickness for the Co nanodisk are $120$ nm and $20$ nm, respectively, with the saturation magnetization $M^\text{Co}_\text{S}=1.4\times 10^6$ A m$^{-1}$ and the exchange constant $A_\text{ex}^\text{Co}=2.5\times 10^{-11}$ J m$^{-1}$. The CoPt nanotrack dimensions are $2000$ nm $\times$ $200$ nm $\times$ $8$ nm, with the saturation magnetization $M_\text{S}^\text{CoPt}=1.4\times 10^6$ A m$^{-1}$, the exchange constant $A_\text{ex}^\text{CoPt}=1.5\times 10^{-11}$ J m$^{-1}$ and the Gilbert damping $\alpha=0.3$. The PMA of the CoPt nanotrack are tested for values varying from $K_{\text{u1}}=2\times 10^{5}$ J m$^{-3}$ to $5\times 10^{5}$ J m$^{-3}$.

\section{Results and Discussion}

\noindent
\textbf{Topological number.}
A magnetic spin texture is characterized by the topological number $Q=\int\rho(x,y)dxdy$ with the topological number density
\begin{equation}
\rho(x,y)=\frac{1}{4\pi}\boldsymbol{m}\cdot(\frac{\partial\boldsymbol{m}}{\partial x}\times\frac{\partial\boldsymbol{m}}{\partial y}),
\label{eq:topological-number}
\end{equation}
where $\boldsymbol{m}$ is the normalized magnetization. The topological number $Q$ of a ground-state skyrmion is strictly equal to $\pm 1$ in the continuous theory. However, due to the lattice structure of the system, we have $Q\approx\pm 1$ for a skyrmion in the micromagnetic system.

\vbox{}\noindent
\textbf{Imprinted spin texture (IST).}
An IST is created in the nanotrack by the interaction between the soft magnetic nanodisk and the nanotrack (see Fig.~\ref{FIG1}). Hysteresis loops for the system with the nanotrack and nanodisk, obtained by applying steps of out-of-plane magnetic field until magnetization achieve the relaxed state, are shown in \blue{Supplementary Figure 1}. The magnetization evolves in the $z$-axis by the interaction between the stable magnetic vortex core and the perpendicular nanotrack magnetization. For anisotropies below $K_{\text{u1}}=4\times 10^5$ J m$^{-3}$, the coercivity increases due to the pinning caused by the vortex core polarization opposite to the nanotrack background magnetization. For higher anisotropy values, the coercivity decreases due to the pinning loss, when the vortex core is aligned with the strong nanotrack perpendicular magnetization.

It should be noted that, just after device fabrication and before imprinting a spin texture, it is necessary to apply perpendicular magnetic field to the sample in order to annihilate the as-grown magnetic domains in the track and saturate it. This is necessary once and for all. It is an intriguing finding that the type of an IST formed depends on the field strength utilized (\blue{see Supplementary Figure 2}).

In Fig.~\ref{FIG2}(a), we show the remanent magnetization recorded under the nanodisk, after applying small magnetic field $(B_{z}=0.15$T) to saturate the magnetization just in the nanotrack. Here, we find that the IST is a skyrmion by calculating the topological number of $Q\approx -1$. The skyrmion diameter decreases with increasing PMA strength.

On the other hand, in Fig.~\ref{FIG2}(b), we show the remanent state of the system after applying large magnetic field $(B_{z}=0.5$T) for the total out-of-plane magnetization saturation; in this case, a chiral configuration could be formed, in which the magnetic vortex core is not opposite anymore to the nanotrack background magnetization. This skyrmion-like structure with the topological number of $Q=0$ is identified as the skyrmionium~\cite{finazzi,zhou2016}, which is composed by a combination of skyrmions with opposite topological numbers, i.e., $Q=+1$ and $Q=-1$. The radius of the skyrmionium does not depends on the PMA strength.

\begin{figure}[t]
\centering
\includegraphics[width=0.50\textwidth]{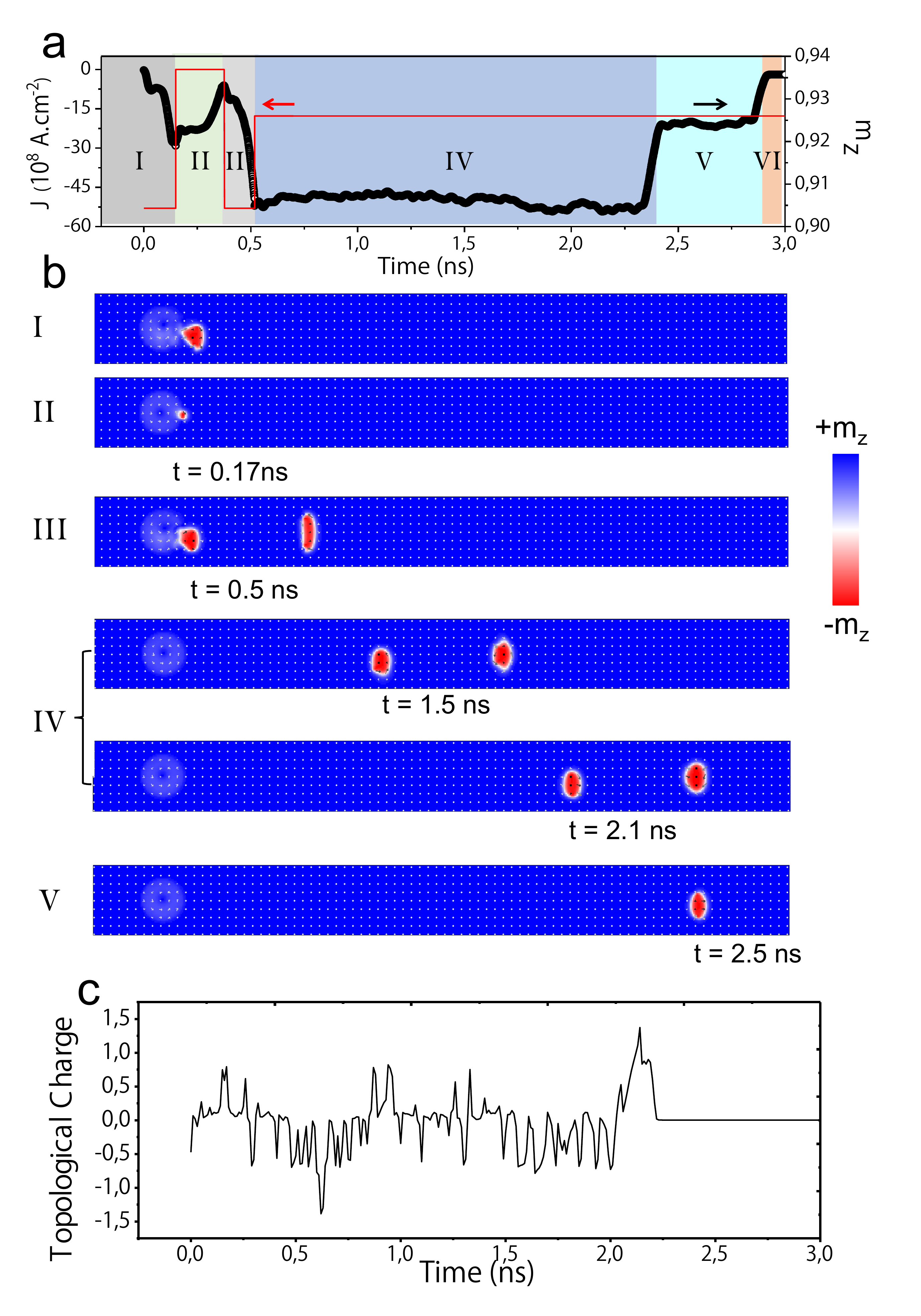}
\caption{(Color online.) Sequential creation and transport of a pair of imprinted RMSs. (a) Time evolution of $m_z$ and $J$ during the creation and transport processes. (b) Top-views of the magnetization in the nanotrack at selected times. (c) Topological number $Q$ of the nanotrack as a function of time. We have used $K_{\text{u1}}=5\times 10^5$ J m$^{-3}$.}
\label{FIG5}
\end{figure}

\vbox{}\noindent
\textbf{Detaching process.}
We apply an uniform and constant spin-polarized current ($J=10\times 10^8$ A cm$^{-2}$ with $p=0.5$) in the nanotrack ($K_{\text{u1}}=3\times 10^5$ J m$^{-3}$) to detach the skyrmion from the underneath of the nanodisk and drive it into motion by way of the spin-transfer torque in the adiabatic approximation~\cite{zang,tatara,iwasaki}, as shown in Fig.~\ref{FIG2}(c) (see also \blue{Supplementary Movie 1}). The mechanism is the same as in the case of the skyrmion transport~\cite{fert1,yu1,nagaosa1}. Higher values of $K_{\text{u1}}$ could be used to decrease the diameter of the moving IST. If the same amount of $J$ is applied for a certain time period, the moving IST rapidly evolves to a domain-wall pair, as shown in Figs.~\ref{FIG2}(c). The topological number of the domain-wall pair becomes large since the spin rotates many times in the region where $m_z=0$.

On the other hand, in the nanotrack with $K_{\text{u1}}=5\times 10^5$ J m$^{-3}$ where a skyrmionium imprinted from the nanodisk is present [Fig.~\ref{FIG2}(b)], a combination of spin-transfer torque and Oersted field from $J=53\times 10^8$ A cm$^{-2}$ is enough to reduce a skyrmion inside the imprinted skyrmionium and expel it aligned with the Oersted field as a magnetic droplet. The droplet rapidly evolves to a small moving magnetic bubble. However, if the same amount of $J$ continues to be applied, it evolves subsequently to a domain wall pair [see Fig.~\ref{FIG2}(d)]. It is necessary to decrease the current $J$ to prevent it from developing into a domain wall pair. If $J$ is weaker than $J_s=18.455\times 10^{8}$ A cm$^{-2}$, it shrinks to a droplet and disappears. The reason is that the IST cannot be stabilized in the absence of the spin-polarized current and the minimum energy takes when the radius is equal to zero. The magnetization $m_z$ becomes constant after the disappearance of the IST. Indeed, a magnetic bubble expands and transformed into a domain wall pair when it touches the edges for the current stronger than $J_s$. Figure~\ref{FIG3} shows the time evolution of the magnetization $m_z$ for various magnitude of the current $J$, where we find that the IST travels straightly when $J=J_s \equiv 18.5\times 10^8$ A cm$^{-2}$. The results show that the spin-polarized current has a force to expand the IST.

\begin{figure*}[t]
\centering
\includegraphics[width=1.00\textwidth]{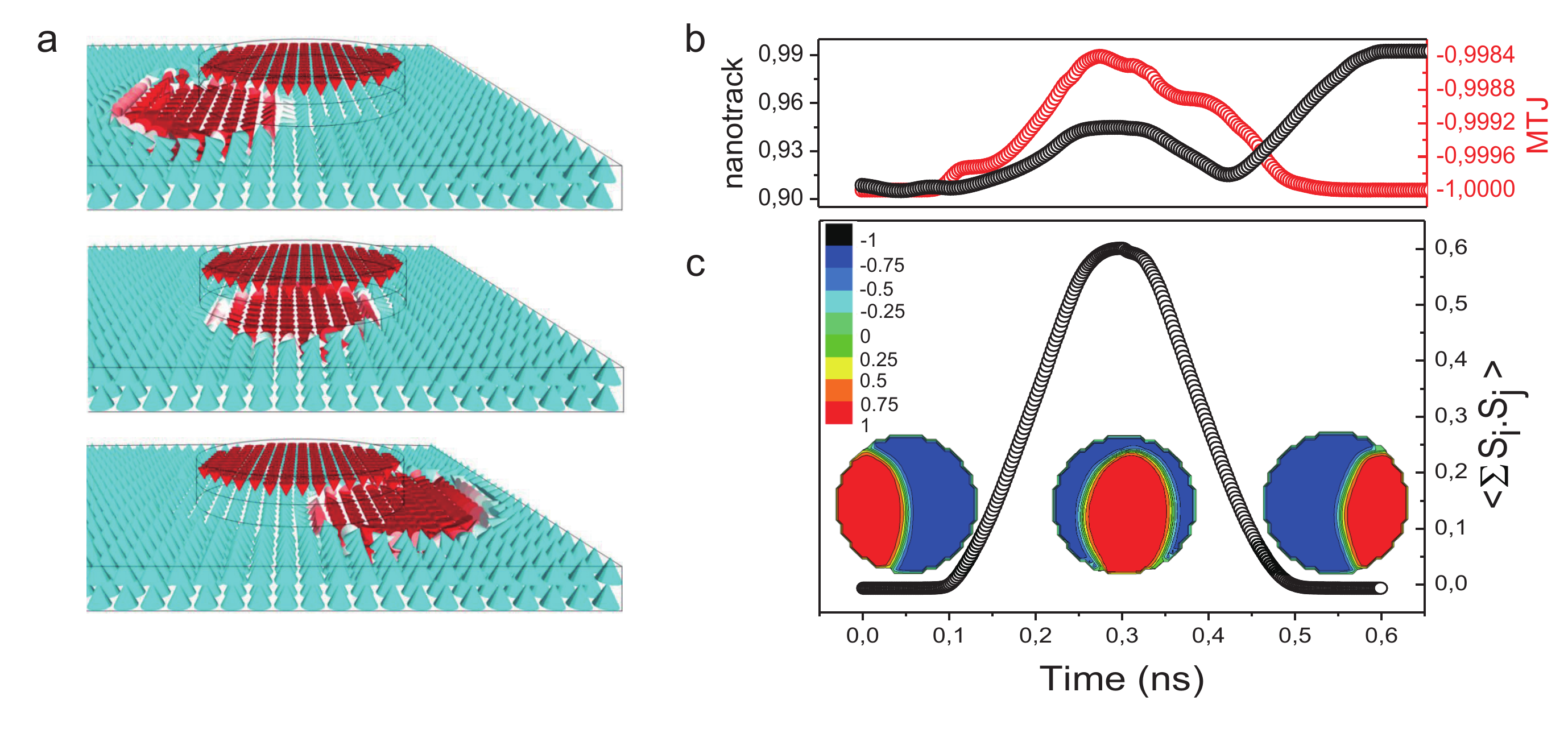}
\caption{(Color online.) (a) Three-dimensional views of the nanotrack and MTJ electrode magnetization during the detection of the moving RMS. We show the cases where the RMS is before, just below, and after the MTJ. (b) Time evolution of $m_z$ for the nanotrack and MTJ. The black curve indicates the magnetization of the nanotrack, while the red curve indicates that of the MTJ. (c) Scalar product between the nanotrack and MTJ spins, showing the tunnel magnetoresistance expected during the measurement. Insets present snapshots of scalar products in different spin texture positions in relation to the MTJ sensor corresponding to the three snapshots.}
\label{FIG6}
\end{figure*}

\begin{figure}[t]
\centering
\includegraphics[width=0.50\textwidth]{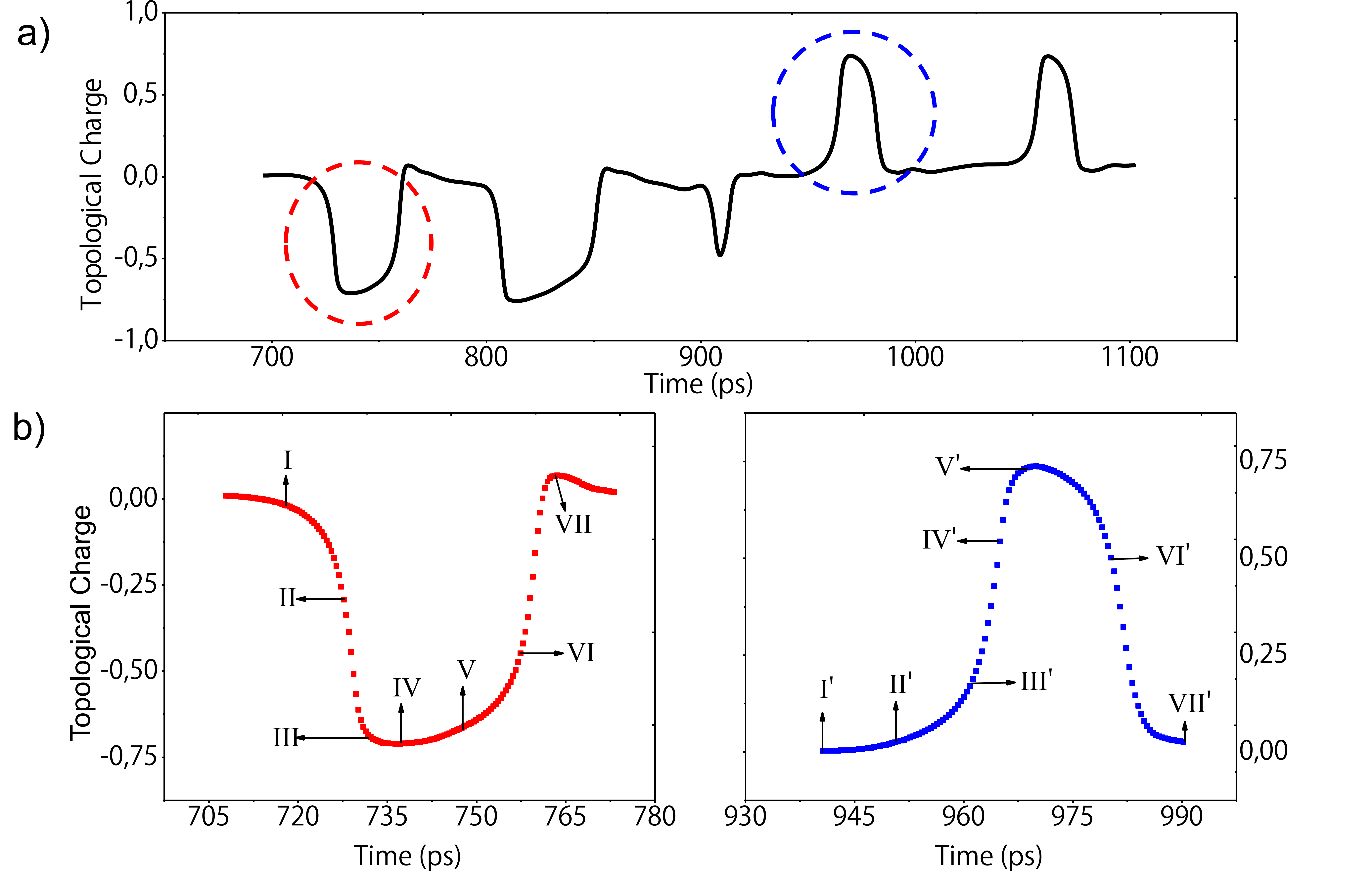}
\caption{(Color online.) (a) Topological number evolution during the transport of a RMS. (b) shows the highlighted peaks of (a) with the minimum and maximum values of $Q$. The spin configuration for each peak are given in Fig.~\ref{FIG.B}.}
\label{FIG.A}
\end{figure}

\begin{figure*}[t]
\centering
\includegraphics[width=1.00\textwidth]{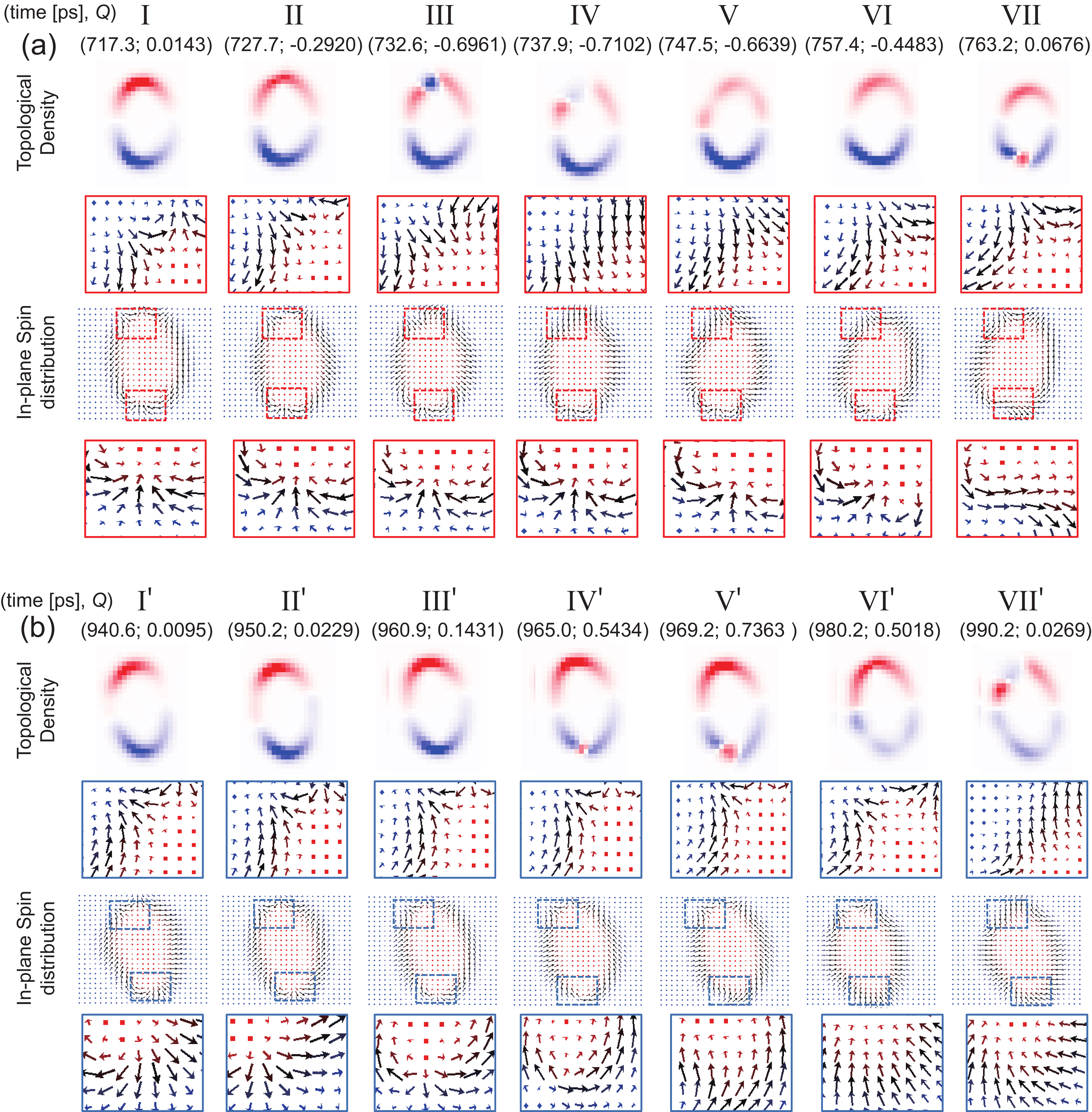}
\caption{(Color online.) Spin configuration and topological number density for the (a) minimum value of $Q$, as shown in Fig.~\ref{FIG.A}(b) and (b) maximum value of $Q$, as shown in Fig.~\ref{FIG.A}(c). The roman numerals corresponds to its respective point in the graph, the time and the spin-component distribution $m_z(x)$ of each state is indexed by ($t$,~$Q$).}
\label{FIG.B}
\end{figure*}

\vbox{}\noindent
\textbf{Transport process.}
To transport a small and stable IST, we utilize fast pulses of $J_\text{c}=53\times 10^8$ A cm$^{-2}$ during $0.17$ ns to create and expel the IST from the underneath of the nanodisk (see \blue{Supplementary Movie 1}). After decreasing the current density to $J_\text{s}=18.5\times 10^8$ A cm$^{-2}$, the IST reaches a stable size of $60$ $\times$ $80$ nm and a steady velocity of $888$ m s$^{-1}$ during the entire way through the $2$-$\mu$m-long nanotrack (see \blue{Supplementary Movie 2}). Small IST deformation occurs due to the strong interaction with the nanodisk magnetization during its creation, similarly to magnetic droplets formation~\cite{zhou2014,zhou2015}, and continues throughout the motion, once there is no external magnetic field applied along the same direction of the IST core to stabilize it, as in the case of the skyrmion investigations performed so far~\cite{romming1,romming2,moreau}.

The evolution of the system magnetization along the $z$-direction ($m_z$) during the creation and the transport of an IST is given in Fig.~\ref{FIG4}. Nearly the stable value of $m_z$ during the transport indicates the size stability, which can also be observed in the snapshots as given in Fig.~\ref{FIG4}(b). Figure~\ref{FIG4}(c) shows the topological number $Q$ of the nanotrack as a function of time during the creation and motion of the IST. It shows that the topological number $Q$ is oscillating, indicating the IST is evolving rapidly between a skyrmion and a bubble. Hence we call the IST the resonant magnetic soliton (RMS) hereafter. Since its average value of the topological number equals zero ($Q=0$), there exists no skyrmion Hall effect, and the RMS moves along a straight line. This is a great advantage for building racetrack-type spintronic devices.

We can create a single dynamical IST with spin-polarized current pulses as described above, and analyze the transport of such spin configuration with stable size under an applied current density of $J_\text{s}=18.5\times 10^{8}$ A cm$^{-2}$. It is important to state that below this current density, the RMS transport slows down until it disappears as shown in Fig.\ref{FIG2}(e) (see also \blue{Supplementary Movie 3}).

\vbox{}\noindent
\textbf{Successive creations of RMSs.}
For a practical utilization of the proposed method in skyrmion-based racetrack memory devices, the density of RMS formation and their susceptibility to dipolar interaction between each other should be tested. For this matter, we apply a new series of current pulses, as presented in Fig.~\ref{FIG5}(a). First, we create a RMS by applying the current $J_\text{c}$ during $0.17$ ns. Just after the creation of the first RMS, the applied current is switched off for $0.22$ ns. This procedure is performed in order to quickly diminish the size of the first RMS, and assure that the successive application of the current $J_\text{c}$ for the formation of a second RMS will restore the size of the first RMS to its original size. By applying this protocol, we find a possible way to create a stable pair with similar size and a small distance between each other ($\sim 300$ nm). Such a process can be observed along the evolution of $m_z$ in Fig.~\ref{FIG5}(a). The motion of the pair was investigated throughout the nanotrack and some snapshots are presented in Fig.~\ref{FIG5}(b) (see \blue{Supplementary Movie 4}). The corresponding topological number $Q$ of the nanotrack is given in Fig~\ref{FIG5}(c).

\vbox{}\noindent
\textbf{Reading process.}
We make use of the MTJ sensor to detect the passing of a RMS through the nanotrack. We have performed numerical calculations for the RMS measurement by a MTJ patterned in the other side of the nanotrack. It is important to investigate the interaction between the RMS and the ferromagnetic electrode, which could pin, deform or destroy these magnetic objects, and estimate tunnel magnetoresistive values expected. We have inserted a CoPt nanodisk with a diameter of $120$ nm and thickness of $4$ nm, and the same $K_{\text{u1}}$ as in the nanotrack, $2$ nm apart to simulate the average thickness of thin oxides used as tunnel barriers in MTJs. As shown in Fig.~\ref{FIG6}(a), we realize that the effect of the RMS interaction with CoPt MTJ electrode is not strong enough to destroy the RMS. As shown in Fig.~\ref{FIG6}(b), we notice a small reduction in the size of the RMS, represented by the positive slight increase in the $m_z$ of the nanotrack. When the RMS is passing below the MTJ electrode, a very small decrease is found in the MTJ electrode $m_z$.

The tunnel magnetoresistance could be estimated as $\langle\sum S_i\cdot S_j\rangle^{-1}$. We calculated scalar products of spins from each cell of both tunnel junction sides, and the results are given in Fig.~\ref{FIG6}(c). Using the same diameter for the MTJ and nanodisk used in the RMS creation ($120$ nm), the highest tunnel magnetoresistance observed is about $60\%$. The insets show the snapshots of scalar products $\sum S_i\cdot S_j$ at selected times, where the red area shows the position of the moving RMS.

\vbox{}\noindent
\textbf{Resonant magnetic soliton (RMS).}
Finally, we explore a detailed mechanism of the fluctuation of the topological number of a RMS. For this purpose, we study the topological number $Q$ as well as the topological number density $\rho(\mathbf{x})$. We show a detailed time-evolution of the topological number in Fig.~\ref{FIG.A}. The topological number fluctuates around $Q=0$ and takes extrema around $Q=\pm 0.75$. In this process, the topological number $|Q|$ suddenly increases (II, III$'$) and reaches a maximum value (IV, V$'$). Then it decreases slowly (V, VI$'$) and returns to be zero (VII, VII$'$). We present snapshots of the topological number density and the corresponding spin configuration in Fig.~\ref{FIG.B}. Especially, the in-plane spin configuration contains a defect-like point at the snapshot time when the topological number takes a maximum value, where the contribution to the topological number is not strictly $Q=\pm 1$. Namely, when a defect-like point is almost formed but not completely, the soliton acquires the topological number $|Q|<1$. We can see that the sudden change of the topological number is induced by the creation and the annihilation of a defect-like point.

The shape of the RMS is elongated perpendicular to the applied current direction and forms an ellipse. The topological number density for the RMS with $Q=0$ is positive for the upper half (indicated by red in Fig.~\ref{FIG.B}), while it is negative for the lower half (indicated by blue in Fig.~\ref{FIG.B}). A defect-like point emerges almost at the top or the bottom of the RMS.

This RMS is different from a traditional magnetic bubble due to the following reasons. First, the size of a RMS is much smaller than that of a traditional magnetic bubble, which is stabilized by the magnetic dipole-dipole interaction and its size is of the order of $\mu$m. The magnetic dipole-dipole interaction does not play a role for the stabilization of a RMS since its size is of the order of $100$ nm. Second, a RMS is stabilized only in the presence of the applied current, while a traditional magnetic bubble is stable in the absence of the applied current. Finally, the topological number of a RMS oscillates as a function of time in a RMS, of which the average value is equal to zero. This is in stark contrast to a moving traditional magnetic bubble, of which the topological number may vary but cannot have an average value of zero (see also \blue{Supplementary Movie 5}).

\section{Conclusion}

We have proposed to create and control a single or successive moving RMSs in a nanotrack without the DMI and without the requirement of an external applied field to ensure the stability. A RMS has the averaged topological number of zero, i.e., $\left\langle Q\right\rangle=0$, and thus is free from the skyrmion Hall effect. It moves straightly along the nanotrack.

As the geometries and materials parameters utilized in this work are similar to the skyrmion imprinting proposed and experimentally demonstrated in Refs.~\onlinecite{sun,miao,gilbert}, this method for the creation of the RMS is expected to work at room temperature. Meanwhile, the advantage of such a method is the increase in the integration density, once there is no need of a third electrode to apply perpendicular current for the spin texture creation as in the most skyrmionic devices proposed so far. The proposed method and device are promising for applications in future racetrack memories with controlled RMS density for creation and high tunnel magnetoresistance signal for detection.

\section*{Acknowledgments}
The authors thank CNPq, CAPES and FAPEMIG (Brazilian agencies) for financial support. X.Z. was supported by JSPS RONPAKU (Dissertation Ph.D.) Program. Y.Z. acknowledges the support by the National Natural Science Foundation of China (Project No. 1157040329) and Shenzhen Fundamental Research Fund under Grant No. JCYJ20160331164412545. M.E. acknowledges the support by the Grants-in-Aid for Scientific Research from JSPS KAKENHI (Grant Nos. 25400317 and 15H05854). This work was also supported by CREST, JST.




\begin{thebibliography}{99}

\bibitem{skyrme} Skyrme. T. H. R., A unified theory of mesons and baryons. \textit{Nucl. Phys.} \textbf{31}, 556-569, (1962)

\bibitem{rossler} Ro{\ss}ler, U. K., Bogdanov, A. N. {\&} Pfleiderer, C. Spontaneous skyrmion ground states in magnetic
metals. \textit{Nature} \textbf{442}, 797-801, (2006).

\bibitem{Muhlbauer} Muhhlbauer, S. \textit{et al.} Skyrmion lattice in a chiral magnet. \textit{Science} \textbf{323}, 915-919, (2009).

\bibitem{neubauer} Neubauer, A. \textit{et al.} Topological Hall effect in the A phase of MnSi. \textit{Phys. Rev. Lett.} \textbf{102}, 186602  (2009).

\bibitem{pappas} Pappas, C. \textit{et al.} Chiral paramagnetic skyrmion-like phase in MnSi. \textit{Phys. Rev. Lett.} \textbf{102}, 197202 (2009).

\bibitem{Yu}  Yu, X. Z. et. al. Real-space observation of a two-dimensional skyrmion crystal. \textit{Nature} \textbf{465}, 901 (2010).

\bibitem{bogdanov} Bogdanov, A. N. {\&} Ro{\ss}ler, U. K. Chiral symmetry breaking in magnetic thin films and multilayers. \textit{Phys. Rev. Lett.} \textbf{87}, 037203 (2001).

\bibitem{bode} Bode, M \textit{et. al.} Chiral magnetic order at surfaces driven by inversion asymmetry. \textit{Nature} \textbf{447}, 190-193 (2007).

\bibitem{dzy} Dzyaloshinsky, I. A thermodynamic theory of weak ferromagnetism of antiferromagnetics. \textit{J. Phys. Chem. Sol.} \textbf{4}, 241-255 (1958).

\bibitem{moriya} Moriya, T. Anisotropic superexchange interaction and weak ferromagnetism. \textit{Phys. Rev.} \textbf{120}, 91-98 (1960).

\bibitem{YuR} Yu, X. Z. et. al. Near room-temperature formation of a skyrmion crystal in thin-films of the helimagnet FeGe. Nat. Mat. \textbf{10}, 106 (2011)

\bibitem{buttner} Buttner, F. \textit{et. al.} Dynamics and inertia of skyrmionic spin structures. \textit{Nat. Phys.} \textbf{11}, 225-228 (2015).

\bibitem{hei} Heinonen, O., Jiang, W., Somaily, H., te Velthuis, S. {\&} Hoffmann, A. Generation of magnetic skyrmion bubbles by inhomogeneous spin Hall currents. \textit{Phys. Rev. B} \textbf{93}, 094407 (2016).

\bibitem{schulz} Schulz, T. \textit{et. al.} Emergent electrodynamics of skyrmions in a chiral magnet. \textit{Nat. Phys.} \textbf{8}, 301-304 (2012).

\bibitem{seki} Seki, S., Yu, X. Z., Ishiwata, S. {\&} Tokura, Y. Observation of skyrmions in a multiferroic material. \textit{Science} \textbf{336}, 198-201 (2012).

\bibitem{parkin1} Parkin, S. S. P., Hayashi, M. {\&} Thomas, L. Magnetic domain-wall racetrack memoriy. \textit{Science} \textbf{320}, 190-194 (2008).

\bibitem{araujo} de Araujo, C. I. L., Alves, S. G., Buda-Prejbeanu, S. D. {\&} Dieny, B. Multilevel thermally assisted magnetoresistive random access memory based on exchange-biased vortex configurations. \textit{Phys. Rev. Appl.} \textbf{6}, 024015 (2016).

\bibitem{tehrani} Tehrani, S., Slaughter, S. M., Chen, E., Durlam, M., Shi, J., {\&} DeHerrera, M. Progress and outlook for MRAM technology. \textit{IEEE Trans. Mag.} \textbf{35}, 2814-2819 (1999). 

\bibitem{par} Parkin, S. S. P. \textit{et. al.} Exchange-biased magnetic tunnel junctions and application to nonvolatile magnetic random access memory. \textit{J. Appl. Phys.} \textbf{85}, 5828 (1999).

\bibitem{yang} Yang, S.-H., Ryu, K.-S.  {\&} Parkin, S. S. P. Domain-wall velocities of up to 750 m s$^{-1}$ driven by exchange-coupling torque in synthetic antiferromagnets. \textit{Nat. Nanotech.} \textbf{10}, 221-226 (2015).

\bibitem{iwasaki} Iwasaki, J., Mochizuki, M. {\&} Nagaosa, N. Universal current-velocity relation of skyrmion motion in chiral magnets. \textit{Nat. Commun.} \textbf{4}, 1463, (2013).

\bibitem{iwasaki2} Iwasaki, J., Mochizuki, M. {\&} Nagaosa, N. Current-induced skyrmion dynamics in constricted geometries. \textit{Nat. Nanotech.} \textbf{8} 742, (2013)

\bibitem{Yin} Yin, G. \textit{et. al.} Topological number analysis of ultrafast single skyrmion creation. \textit{Phy. Rev. B} \textbf{93}, 174403 (2016). 

\bibitem{silva} Silva, R. L., Secchin, L. D., Moura-Melo, W. A., Pereira, A. R. {\&} Stamps, R. L. Emergence of skyrmion lattices and bimerons in chiral magnetic thin films with nonmagnetic impurities. \textit{ Phys. Rev. B} \textbf {89}, 05443 (2014).

\bibitem{sampaio} Sampaio, J., Cros, V., Rohart, S., Thiaville, A. {\&} Fert, A. Nucleation, stability and current-induced motion of isolated magnetic skyrmions in nanostructures. \textit{Nat. Nanotech.} \textbf{8}, 839-844 (2013).

\bibitem{zhang} Zhang, X. \textit{et. al.} Skyrmion-skyrmion and skyrmion-edge repulsions in skyrmion-based racetrack memory. \textit{Scientific Reports.} \textbf{5}, 7643 (2015).

\bibitem{zhang9400} Zhang, X., Ezawa, M., {\&} Zhou, Y. Magnetic skyrmion logic gates: conversion, duplication and merging of skyrmions. \textit{Sci. Rep.} \textbf{5}, 9400, (2015).

\bibitem{zhangV} Zhang, X. \textit{et. al.} Magnetic skyrmion transistor: skyrmion motion in a voltage-gated nanotrack. \textit{Sci. Rep.} \textbf{5}, 11369, (2015).

\bibitem{kang} Kang, W., \textit{et. al.} Complementary skyrmion racetrack memory with voltage manipulation. \textit{IEEE Electron Device Lett.} \textbf {37}, 924-927 (2016).


\bibitem{xing1} Xing, X., Pong, P. W. T., {\&} Zhou, Y. Skyrmion domain wall collision and domain wall-gated skyrmion logic. \textit{Phys. Rev. B} \textbf {94}, 05443 (2016). 

\bibitem{woo} Woo, S. \textit{et. al.} Observation of room-temperature magnetic skyrmions and their current-driven dynamics in ultrathin metallic ferromagnets. \textit{Nat. Mater.} \textbf{ 15}, 501-506 (2016).

\bibitem{jiang} Jiang, W. \textit{et. al.} Blowing magnetic skyrmion bubbles. \textit{Science} \textbf{349}, 283-286, (2015).


\bibitem{sun} Sun, L. \textit{et. al.} Creating an artificial two-dimensional skyrmion crystal by nanopatterning. \textit{Phys. Rev. Lett.} \textbf{110}, 167201, (2013).

\bibitem{miao} Miao, B. F. \textit{et. al.} Experimental realization of two-dimensional artificial skyrmion crystals at room temperature. \textit{Phys. Rev. B} \textbf{90}, 174411, (2014).

\bibitem{gilbert} Gilbert, D. A. \textit{et. al.} Realization of ground-state artificial skyrmion lattices at room temperature. \textit{Nat. Commun.} \textbf{6}, 8462, (2015).

\bibitem{finazzi}  Finazzi, M. \textit{et. al.}  Laser-induced magnetic nanostructures with tunable topological properties. 
\textit{Phys. Rev. Lett.} \textbf{110}, 177205 (2013)

\bibitem{zhou2016} Zhang, X. \textit{et. al.} Control and manipulation of a magnetic skyrmionium in nanostructures. \textit{Phys. Rev. B} \textbf{94}, 094420 (2016).

\bibitem{zhou2014} Zhou, Y. {\&} Ezawa, M. A reversible conversion between a skyrmion and a domain-wall pair in a junction geometry. \textit{Nat. Commun.} \textbf{5}, 4652 (2014).

\bibitem{zhou2015} Zhou, Y. \textit{et. al.}. Dynamically stabilized magnetic skyrmions. \textit{Nat. Commun.}. \textbf{6}, 8193 (2015).

\bibitem{mumax} Vansteenkiste, A. \textit{et. al.} The design and verification of MuMax3.\textit{AIP Adv.} \textbf{4}, 107133 (2014).

\bibitem{zang} Zang, J., Mostovoy, M., Han, J. H. {\&} Nagaosa, N. Dynamics of skyrmion crystals in metallic thin films. \textit{Phys. Rev. Lett.} \textbf{107}, 136804, (2011).

\bibitem{tatara} Tatara, G. \textit{et. al.} Threshold current of domain wall motion under extrinsic pinning, $\beta$-term and non-adiabaticity. \textit{J. Phys. Soc. Japan} \textbf{75}. 064708 (2006).



\bibitem{fert1} Fert, A., Cros, V. {\& } Sampaio, J. Skyrmions on the track. \textit{Nat. Nanotech.} \textbf{8}, 152-156 (2013).

\bibitem{yu1} Yu, X. Z., \textit{et. al.} Skyrmion flow near room temperature in an ultralow current density. \textit{Nat. Commun.} \textbf{3}, 988, (2013)

\bibitem{nagaosa1} Nagaosa, N. {\&} Tokura, Y. Topological properties and dynamics of magnetic skyrmions. \textit{Nat. Nanotech.} \textbf{8}, 899-911 (2013).

\bibitem{romming1} Romming, N. \textit{et. al.} Writing and deleting single magnetic skyrmions. \textit{Science} \textbf{341}, 636-639 (2013).

\bibitem{romming2} Romming, N., Kubetzka, A., Hanneken, C., von Bergmann, K. {\&} Wiesendanger, R. Field-dependent size and shape of single magnetic skyrmions. \textit{Phys. Rev. Lett.} \textbf{114}, 177203, (2015).

\bibitem{moreau} Moreau-Luchaire, C. \textit{et. al.} Additive interfacial chiral interaction in multilayers for stabilization of small individual skyrmions at room temperature.\textit{Nat. Nanotech.} \textbf{11}, 444-448 (2016).


\end{thebibliography}
\end{document}